\documentclass[superscriptaddress,twocolumn,nofootinbib]{revtex4}
\usepackage{amssymb, bm}
\usepackage{graphicx}
\usepackage{hyperref}

\begin{document}

\title{Effect of spin-orbit coupling on spectral and
transport properties of tubular electron gas in InAs nanowires}

\author{I.A. Kokurin}
\email{kokurinia@math.mrsu.ru} \affiliation{A. F. Ioffe
Physical-Technical Institute, Russian Academy of Sciences, 194021
St. Petersburg, Russia} \affiliation{Institute of Physics and
Chemistry, Mordovia State University, 430005 Saransk, Russia}

\date{\today}

\begin{abstract}
We constructed the Hamiltonian of spin-orbit splitting for carriers
of a tubular electron gas in InAs nanowires. The spectral problem is
solved using an exact numerical diagonalization. It is shown that
the contribution of $k$-linear Dresselhaus-like spin-orbit (SO)
coupling leads to renormalization of so-called SO-gaps and
appearance of anticrossings in subband spectrum. These features can
be detected in ballistic transport.
\end{abstract}


\maketitle

\section{Introduction}
\label{sec:Intro}

Nanowires (NWs) of narrow gap III-V semiconductors (such as InAs)
attract significant interest in the field of modern nanoelectronics.
NW is a good candidate for application in nanodevices such as field
effect transistor (FET)~\cite{Dayeh2007,Chuang2013}. The
near-surface band bending and Fermi-level pinning lead to formation
of a two-dimensional electron gas (2DEG) close to the surface of an
InAs NW~\cite{Hernandez2010}. Thus, a one-dimensional (1D) tubular
conducting channel arises near the NW surface (see
Fig.~\ref{fig1}a). Furthermore, the asymmetric confinement of such
an tubular electron gas (TEG) leads to strong spin-orbit coupling
(SOC) of Rashba type~\cite{Bychkov1984}. A possibility of SOC
strength tuning by gates of different geometry~\cite{Liang2012}
allows to utilize the NWs in spintronics, e.g. as the basic element
of Datta-Das spin-FET~\cite{Datta1990} or a gate-defined spin-orbit
qubit \cite{Nadj-Perge2010}.

Ballistic transport is preferable for spintronic nanodevices but
implementation of the ballistic transport regime in InAs NWs has
long hampered due to low carrier mobility that is determined by the
surface roughness scattering. Nevertheless, it was recently shown
that the ballistic transport can be realized in short InAs
NWs~\cite{Chuang2013}.

The effect of Rashba SOC (RSOC) on the energy spectrum and ballistic
transport in InAs NW was recently studied \cite{Kokurin2014}.
However, the influence of the lack of inversion center in
semiconductor material constituting the nanostructure
\cite{Dyakonov1986} was not considered. The so-called $k$-linear
Dresselhaus SOC (DSOC) \cite{Dyakonov1986} arises from the following
contribution to the Hamiltonian of $\Gamma_6$ conduction band of
bulk III-V semiconductor \cite{Dresselhaus1955}

\begin{equation}
\label{cubic} H_\mathrm{D}=\gamma\bm{\sigma\kappa},
\end{equation}
where $\gamma$ is the bulk Dresselhaus parameter and vector
$\kappa=(\kappa_x,\kappa_y,\kappa_z)$ has components
$\kappa_x=k_x(k_y^2-k_z^2)$ and $\kappa_y$, $\kappa_z$ can be
obtained by cyclic permutations, and
$\bm\sigma=(\sigma_x,\sigma_y,\sigma_z)$ is the vector of Pauli
matrices. It should be noted that this operator is written in
principle axes, i.e. $x||[100]$ etc. It is obvious that dimension
lowering (to 2D or 1D) results in orientation-dependent SOC. NWs
usually grows at orientation along [111]-direction. The specificity
of TEG is that electrons undergo different DSOC in each point of NW
cross-section in contrast to usual planar
structures~\cite{Zhang2006,Wang2009}.

In present work we study the effect of SOC on spectral properties of
NW using the simple model of 2DEG placed on a cylindrical surface
\cite{Magarill1996}. The ballistic transport (conductance and
thermopower) is studied as well.

\section{Hamiltonian and spectral problem}

\begin{figure}
\includegraphics[width=\columnwidth]{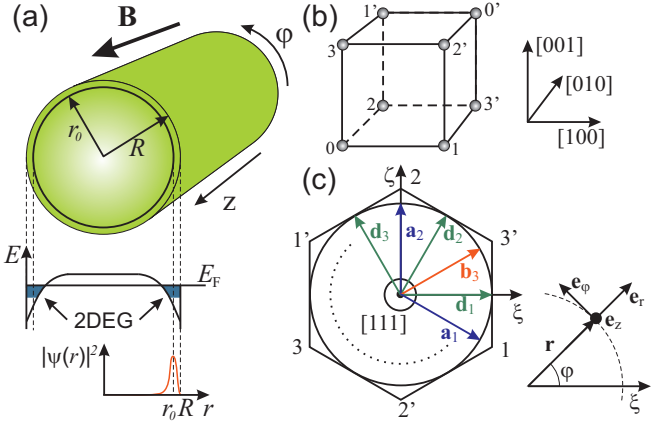}
\caption{\label{fig1} (Color online) (a) Sketch of the InAs NW with
radius $R$ placed in longitudinal magnetic field $\mathbf{B}$. The
band bending which leads to TEG formation is depicted. Carriers are
concentrated in thin cylindrical layer of radius $r_0$. The radial
distribution is schematically shown. (b) The cubic cell of a
zinc-blende lattice. Atoms of one sort (e.g. In) are shown in
vertices of the cube. (c) The scheme that is needed for DSOC
Hamiltonian derivation. The sets of points $1$, $2$, $3$ ($1'$,
$2'$, $3'$) are projected on (111)-plane. The NW axis coincides with
[111]-direction. Other vectors lie in (111)-plane. Any unit vector
in this plane is determined by only one parameter (angle $\varphi$),
see Appendix~\ref{sec:AppA}. The polar frame of reference in plane
(111) and unit vectors discussed in text are shown as well.}
\end{figure}

The [111]-grown NWs as a rule have a hexagonal cut. However, the
most authors do not take into account this fact and suppose the NW
cross-section to be circular
\cite{Hernandez2010,Jin2010,Bringer2011}. Here we use the simple
model of 2DEG placed on cylindrical surface \cite{Magarill1996}. The
possibility of this model application to TEG of radius $r_0$ (see
Fig.~\ref{fig1}a) in InAs NW was discussed in
Ref.~\cite{Kokurin2014}. The Hamiltonian of the system can be
written in the form $H=H_k+H_\mathrm{R}+H_\mathrm{D}$, where $H_k$
is the kinetic term and $H_\mathrm{R(D)}$ is the operator of RSOC
(DSOC). If the magnetic field $\mathbf{B}$ is included in
consideration then the Zeemann splitting $H_\mathrm{Z}$ has to be
taken into account and we have to include the influence of magnetic
field on orbital motion by substitution $\mathbf{p}\rightarrow
\mathbf{p}-(e/c)\mathbf{A}$ with $\mathbf{p}$ and $\mathbf{A}$ being
the electron momentum and vector potential, respectively. In this
case the Hamiltonian of the system with RSOC only
$H_0=H_k+H_\mathrm{R}+H_\mathrm{Z}$ is given by \cite{Kokurin2014}
\begin{equation}
\label{H_0}
H_0=\frac{\Pi_z^2+\Pi_\varphi^2}{2m}+\frac{\alpha}{\hbar}
(\sigma_z\Pi_\varphi-\sigma_\varphi
\Pi_z)+\frac{1}{2}g^*\mu_\mathrm{B}B\sigma_z,
\end{equation}
where $m$, $g^*$, $\mu_\mathrm{B} =|e|\hbar/2m_0c$ are the effective
mass, $g$-factor and Bohr magneton, respectively,
$\bm{\Pi}=\mathbf{p}-\frac{e}{c}\mathbf{A}$ is the kinematic
momentum and $p_z=-i\hbar\partial/\partial z$,
$p_\varphi=-(i\hbar/r_0)\partial/\partial\varphi$. We use the
following gauge of vector potential, $(A_\varphi=Br_0/2,A_z=0)$.
Here we use the `cylindrical' Pauli matrices
$\sigma_r(\varphi)=\cos\varphi\sigma_x+\sin\varphi\sigma_y$,
$\sigma_\varphi(\varphi)=-\sin\varphi\sigma_x+\cos\varphi\sigma_y$.
The explicit form of $\sigma_{r(\varphi)}$ is
\begin{equation}
\sigma_r(\varphi)=\left(\begin{array}{cc}
0& e^{-i\varphi}\\
e^{i\varphi}&0
\end{array}\right),\qquad
\sigma_\varphi(\varphi)=\left(\begin{array}{cc}
0& -ie^{-i\varphi}\\
ie^{i\varphi}&0
\end{array}\right).
\end{equation}

The spectrum of Hamiltonian (\ref{H_0}) is given by

\begin{eqnarray} \label{spectrum} \nonumber
\frac{E_{jm}(k)}{E_0}=(kr_0)^2+\left(j+\Phi/\Phi_0\right)^2+1/4-\Lambda_\mathrm{R}/2\\
+s\sqrt{(\Lambda_\mathrm{R}
kr_0)^2+[\Delta-(1-\Lambda_\mathrm{R})(j+\Phi/\Phi_0)]^2},
\end{eqnarray}
where $E_0=\hbar^2/2mr^2_0$ is the character energy scale in the
problem, $\Lambda_\mathrm{R}=2m\alpha r_0/\hbar^2$ is the
dimensionless RSOC parameter, and $j=\pm 1/2, \pm 3/2,...$ is
$z$-projection of total angular momentum, $s=\pm 1$ numerates two
branches of spin-splitted spectrum, $\Phi$ is the magnetic flux $\pi
r^2_0B$ through the section of TEG, $\Phi_0=2\pi\hbar c/|e|$ is the
flux quantum, and $2\Delta=g^*\mu_\mathrm{B}B/E_0$ is the
dimensionless energy of Zeemann splitting.

Eigenstates of $H_0$ are given by
\begin{equation}
\label{eigenstates} \Psi(\varphi, z)=\frac{e^{ikz}}{\sqrt{2\pi L}}
\left(\matrix{e^{i(j-1/2)\varphi}C_{jk}^{(s)}\cr
e^{i(j+1/2)\varphi}D_{jk}^{(s)}\cr}\right),
\end{equation}
where $L$ is NW length, $\hbar k$ longitudinal momentum, and spinor
components $C_{jk}^{(s)}$ and $D_{jk}^{(s)}$ are in general
$k$-depndent due to SOC. Normalized eigenspinors are given by
\begin{equation}
\label{spinors}
C_{jk}^{(+)}=D_{jk}^{(-)}=i\sin\left(\frac{\theta_{jk}}{2}\right),\;\;
D_{jk}^{(+)}=C_{jk}^{(-)}=\cos\left(\frac{\theta_{jk}}{2}\right),
\end{equation}
with
\begin{equation}
\theta_{jk}=\arctan\left[\frac{\Lambda_\mathrm{R}
kr_0}{(1-\Lambda_\mathrm{R})\lambda_j-\Delta}\right]
+\pi\theta[\Delta-(1-\Lambda_\mathrm{R})\lambda_j],
\end{equation}
where $\theta(x)$ is the Heaviside unit step function and
$\lambda_j=j+\Phi/\Phi_0$.

As was shown in Ref.~\cite{Kokurin2014} the strong SOC leads to
appearance of so-called W-shape subbands (subbands with $s=-1$ can
have 2 minima and 1 maximum). In this case the so-called SO-gaps
first mentioned in Ref.~\cite{Quay2010} take place in the subband
spectrum. This is the gap between maxima in W-shape subband and
minima in in higher lying subband (with single extremum) that is due
to SOC.

Now we consider the effect of bulk inversion asymmetry (BIA) or
so-called Dresselhaus effect \cite{Dresselhaus1955} that take place
in semiconductors without inversion center, such as InAs. In the
used model of TEG for NW oriented along [111]-axis the $k$-linear
Dresselhaus type spin-orbit Hamiltonian has the form
\begin{eqnarray}
\label{H_D} \nonumber H_\mathrm{D}=\frac{\beta}{\sqrt
6}\left[-\frac{1}{\sqrt
2}\left(\sigma_r(\varphi)k_\varphi-\frac{i}{2r_0}\sigma_\varphi(\varphi)\right)+\sigma_r(-2\varphi)k_z\right.\\
+\left.3\sigma_z\left(\sin3\varphi k_\varphi-\frac{3i}{2r_0}\cos
3\varphi\right)\right],
\end{eqnarray}
where $\beta$ is the $k$-linear DSOC parameter (or BIA parameter),
and $\hbar\mathbf{k}=\mathbf{p}$. Derivation of this operator is
presented in Appendix~\ref{sec:AppA}. The effect of magnetic field
on operator $H_\mathrm{D}$ will be taken into account in standard
manner. Using the above-mentioned gauge of vector potential it leads
to the substitution $k_\varphi\rightarrow
K_\varphi=k_\varphi+\frac{1}{r_0}\frac{\Phi}{\Phi_0}$.

One can see that operator $H_\mathrm{D}$ (\ref{H_D}) and in turn the
total Hamiltonian $H=H_0+H_\mathrm{D}$ do not commute with the
operator of $z$-projection of total angular momentum
$j_z=-i\hbar\frac{\partial}{\partial\varphi}+\frac{\hbar}{2}\sigma_z$
(the rotational invariance of the total Hamiltonian $H$ is broken
because DSOC depends on crystallographic orientation), but
$H_\mathrm{D}$ commutes with $k_z$ (the translational invariance is
conserved) that simplifies the following numerical diagonalization.
This means that the states (\ref{eigenstates}) which had definite
$j$ now will be mixed and anticrossings will appear in the energy
spectrum.

Now we use $20\times 20$ Hamiltonian (the matrix elements are
calculated in basis of states (\ref{eigenstates}) with $j=\pm
1/2,...,\pm 9/2$ and $s=\pm 1$) for numerical diagonalization that
ensure a perfect precision for the spectrum of first 10 low-lying
subbands. The result of numerical diagonalization is depicted in
Fig.~\ref{fig2}. The matrix elements of $H_\mathrm{D}$, that we use,
are written in Appendix~\ref{sec:AppB}. The second and third terms
of Eq.~(\ref{H_D}) mixes the states of subbands $(j,s)$ and $(j\pm
3,\pm s)$. One can see in Fig.~\ref{fig2} the anticrossing that is
due to mixing of states $(j=-3/2,s=+1)$ and $(j=3/2,s=-1)$ by DSOC
Hamiltonian. Close to this anticrossing the energy spectrum can be
described in more simple manner. In this case one can use the
conventional perturbation theory for degenerate states and solve the
secular equation for $2\times 2$ Hamiltonian. However, the form of
matrix elements (see Appendix~\ref{sec:AppB}) does not allow to
obtain a simple analytical result.

\begin{figure}
\includegraphics[width=\columnwidth]{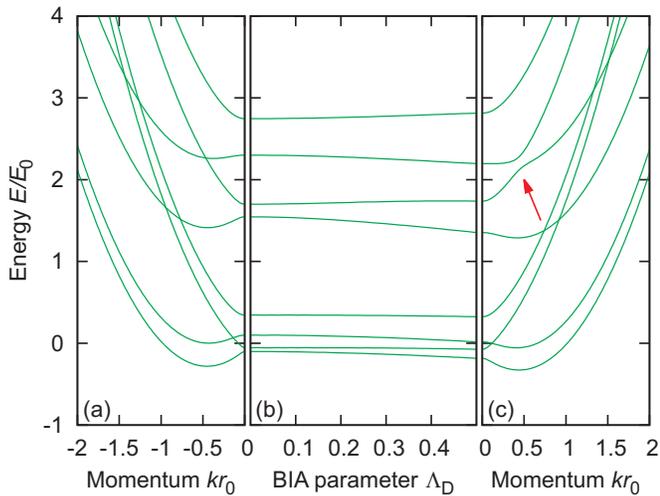}
\caption{\label{fig2} (Color online) The effect of BIA SOC on the
energy spectrum of InAs NW. (a) The subband energy spectrum.
$\Lambda_\mathrm{R}=0.9$, $\Lambda_\mathrm{D}=0$, $\Phi=0.15\Phi_0$.
(b) The dependence of energy at $k=0$ on dimensionless BIA parameter
$\Lambda_\mathrm{D}$. (c) The subband energy spectrum.
$\Lambda_\mathrm{R}=0.9$, $\Lambda_\mathrm{D}=0.5$,
$\Phi=0.15\Phi_0$. An arrow indicates the anticrossing that is due
to BIA SOC.}
\end{figure}

The DSOC Hamiltonian due to the first term in Eq.(\ref{H_D})
increases so-called SO-gaps in spectrum that one can see in
Fig.~\ref{fig2}b. The first term in Hamiltonian (\ref{H_D}) commutes
with $j_z$-operator and has non-zero matrix elements (in basis
(\ref{eigenstates})) only for $j=j'$ (see Appendix~\ref{sec:AppB}).
Thus, the spectral problem for reduced total Hamiltonian having only
first term from $H_\mathrm{D}$ can be solved analytically (as in
Ref.~\cite{Kokurin2014}). This leads to renormalization of SO-gaps
width (in units of $E_0$)
\begin{equation}
\label{SO_gap}
\delta_j^{so}=2\sqrt{[(1-\Lambda_\mathrm{R})\lambda_j-\Delta]^2+\Lambda_\mathrm{D}^2\lambda_j^2/12},
\end{equation}
that tends to Eq.~(6) of Ref.~\cite{Kokurin2014} at
$\Lambda_\mathrm{D}=2m\beta r_0/\hbar^2\rightarrow 0$. However,
taking into account the total $H_\mathrm{D}$ Hamiltonian,
Eq.~(\ref{SO_gap}) is valid far from anticrossings (that are due to
second and third term in Eq.~(\ref{H_D})). When $\Lambda_\mathrm{D}$
increases up to 1 or more then the spectrum varies considerably.
However, we suppose that in real InAs NW the condition $\alpha
>\beta$ ($\Lambda_\mathrm{R}>\Lambda_\mathrm{D}$) fulfills by analogy with planar InAs
nanostructures \cite{Park2013}.

The energy distance between single maximum and two minima in W-shape
subband is given by (using simplified Hamiltonian
$H'=H_0+H_{\mathrm{D}1}$)

\begin{equation}
\label{delta_w} \delta^w_j=\frac{1}{\Lambda_{\mathrm
R}^2}\left(\frac{\Lambda_{\mathrm
R}^2}{2}-\sqrt{[(1-\Lambda_\mathrm{R})\lambda_j-\Delta]^2+\Lambda_\mathrm{D}^2\lambda_j^2/12}\right)^2.
\end{equation}

Let us note that $(j,s=-1)$-subband has W-shape form when the
expression in brackets of Eq.~(\ref{delta_w}) is positive. One can
see from this equation the competition between RSOC and DSOC, i.e.
the increasing of $\Lambda_\mathrm{R}$ leads to appearance of
W-shape subbands, whereas the increasing of $\Lambda_\mathrm{D}$
suppress the emergence of such subbands. The increasing of
$\delta^{so}_j$ and decreasing of $\delta^w_j$ is seen in
Fig.~\ref{fig2} when BIA parameter $\Lambda_\mathrm{D}$ grows up. In
next section we will see that the mentioned SO-gaps or their
renormalization and disappearance of W-shape subbands can be
detected in ballistic transport, namely in conductance and
thermopower.

\section{Ballistic transport}

If the 1D system have the complex spectrum (more than one extremum
in any subband) then the ballistic conductance is given by
\cite{Pershin2004}

\begin{equation}
\label{conductance} G=\frac{G_0}{2}\sum_{in}\beta_i^n
f(E_i^n,\mu,T).
\end{equation}
Here $G_0=e^2/\pi\hbar$ is the conductance quantum (for
spin-degenerate case), $f(E,\mu,T)$ is the Fermi distribution
function, $\mu$ and $T$ are the chemical potential and temperature,
respectively, $E_i^n$ is the energy at $n$-th extremum of $i$-th
subband, and $\beta_i^n=+1$ if $n$-th extremum of $i$-th subband is
the minimum point but $\beta_i^n=-1$ if $n$-th extremum of $i$-th
subband is the maximum one. The sum in (\ref{conductance}) is over
all extremal points of all subbands.

\begin{figure}
\begin{center}
\includegraphics[width=0.75\columnwidth]{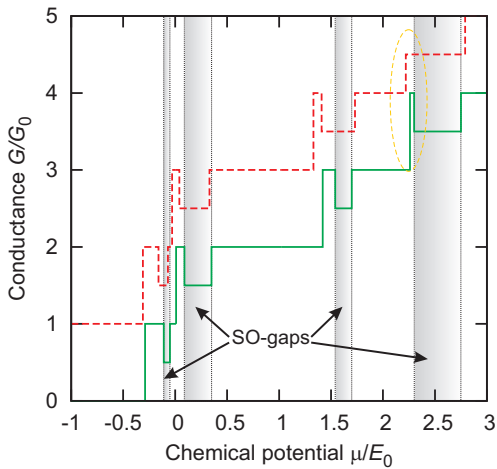}
\caption{\label{fig3} (Color online) Conductance of InAs NW as a
function of the chemical potential. $\Lambda_\mathrm{R}=0.9$,
$\Phi=0.15\Phi_0$, $T=0$; full line, $\Lambda_\mathrm{D}=0$; dashed
line, $\Lambda_\mathrm{D}=0.4$. Curves are vertically shifted for
clarity. The areas corresponding to $\mu$-values lying in SO-gaps
are shaded.}
\end{center}
\end{figure}

The dependence of NW conductance on the chemical potential at $T=0$
is plotted in Fig.~\ref{fig3}. One can see that conductance
step-like behavior differs from one for 1D systems with simple
parabolic dispersion when the conductance always increase with the
chemical potential growing (see for instance
Refs.~\cite{Buttiker1990,Bogachek1993}). It is obvious that
$\sqcup$-like plateaus are due to presence of consequent maxima and
minima in the subband spectrum and the width of such a plateau is
equal to corresponding SO-gap width $\delta^{so}_j$. The comparison
for the case of $\beta=0$ and $\beta\neq 0$ is performed. One can
see that at finite $\Lambda_\mathrm{D}$ there is an distortion of
the picture at $\Lambda_\mathrm{D}=0$: the width of SO-gaps grows up
at $\Lambda_\mathrm{D}$ increasing in accordance with
Eq.~(\ref{SO_gap}). However, at high $\beta$
($\Lambda_\mathrm{D}\sim 1$) the dependence $G(\mu)$ is
significantly distorted. The temperature increasing leads to the
smearing of conductance steps, since electrons coming in from
reservoirs no longer have a sharp step-like energy distribution.

The contribution of W-shape subband, where 2 minima give $+G_0$ and
a maximum gives $-G_0/2$ in accordance with Eq.~(\ref{conductance}),
can be interpreted as a contribution of one electron and one hole
subband. In this sense, if the chemical potential lies between
minima and maximum then we have both electron and hole contribution
$G_0/2+G_0/2=G_0$, and when the chemical potential lies higher than
single maximum then we have only electron contribution $+G_0/2$.
Thus, the $\sqcap$-like plateau deals with a W-shape subband and its
width is equal to $\delta^w_j$ (see Eq.~(\ref{delta_w})). In
Fig.~\ref{fig3} close to $\mu=2.2E_0$ one can see that finite
$\Lambda_\mathrm{D}$ leads to disappearance of the W-like shape of
one subband.

For the thermopower $S$ we recently found the following expression
\cite{Kokurin2014}
\begin{equation}
\label{thermopower} S=\frac{k_\mathrm{B}}{e} \frac{\sum_i\left[\ln
2+\sum_n\beta_i^n
F\left(\frac{E_i^n-\mu}{2T}\right)\right]}{\sum_{in}\beta_i^n
f(E_i^n,\mu,T)},
\end{equation}
where the function $F(x)=\ln(\cosh x)-x\tanh x$ has the following
properties: $F(-x)=F(x)$, $F(\pm\infty)=-\ln 2$, and $F(0)=0$. The
function $F[(E-\mu)/2T]$ as a function of the chemical potential
$\mu$ represents a narrow symmetric peak with a width about several
$T$.

\begin{figure}
\begin{center}
\includegraphics[width=0.73\columnwidth]{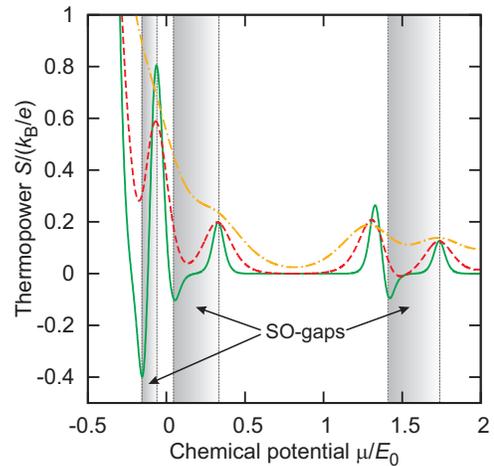}
\caption{\label{fig4} (Color online) Thermopower of InAs NW as a
function of the chemical potential. $\Lambda_\mathrm{R}=0.9$,
$\Lambda_\mathrm{D}=0.4$, $\Phi=0.15\Phi_0$; full line, $T=0.02E_0$;
dashed line, $T=0.05E_0$; dot-dashed line, $T=0.1E_0$. The areas
corresponding to $\mu$-values lying in SO-gaps are shaded.}
\end{center}
\end{figure}

The NW thermopower as a function of the chemical potential is
plotted in Fig.~\ref{fig4}. One can see the negative dips that is
due to maxima in subband spectrum and it is unusual for systems with
a simple parabolic dispersion
\cite{Streda1989,Proetto1991,Kokurin2004}. By analogy with the
conductance quantization these dips can be treated using the hole
representation. It is worth noting that the peaks and dips take
place at $\mu$-values coinciding with energy extrema only
approximately. The numerator in Eq.~(\ref{thermopower}) has extrema
at $\mu=E_i^n$, but in the vicinity of these points the denominator
(dimensionless conductance) varies its own magnitude monotonically
about $1$ that leads to the shift of peak (dip) position and weak
asymmetry of the peak (dip). Moreover, the mentioned deviation
depends on the temperature: the position of peaks (dips) do not
coincide with $\mu=E_i^n$ at high temperatures when the peaks (dips)
overlap each other. Thus, the distance between consequent dip and
peak approximately coincides with corresponding SO-gap (see
Fig.~\ref{fig4}). It is worth noted that the temperature increasing
leads to a disappearance of dips because of the widening of
neighboring peaks of higher amplitudes.

At low temperatures ($T$ is much less than energy distance between
neighboring extrema) Eq.(\ref{thermopower}) is well approximated by
so-called Mott formula \cite{Cutler1969} that connects the
thermopower with conductance.

\section{Conclusion}

Additionally to RSOC we constructed the DSOC Hamiltonian for TEG in
[111]-oriented InAs NW and numerically solved the spectral problem.
It is shown that DSOC lead to appearance of any specific features in
the 1D subband spectrum, such as anticrossings, that was absent at
RSOC consideration only. The effect of longitudinal magnetic field
is also studied. The competition between two types of SOC was
discussed. The ballistic conductance and thermopower of InAs NW were
studied as well. At strong SOC the conductance step-like behavior
differs from one for 1D systems with a simple parabolic dispersion,
that is due to the presence of maxima in subband spectrum that in
turn deals with strong SOC: the conductance can decrease with the
Fermi energy increasing. Such an behavior of $G(\mu)$-dependence
looks like one that take place in semimetallic NW, when subbands of
valence band and conduction one overlap one another, i.e. two types
of charge carrier (electrons and holes) participate in the
conduction process. The thermopower features are the negative dips
in $S(\mu)$ dependence, i.e. $S$ can change the sign with the
chemical potential increasing. The reason of these features
emergence is the same as for conductance decreasing and is due to
strong SOC.

\section*{Acknowledgement}

The author is grateful to N.S. Averkiev, A.M. Monakhov and L.E.
Golub for useful discussions. This work has been supported by
Russian Ministry of Education and Science (project No.~2665).

\appendix

\section{Derivation of $k$-linear Dresselhaus Hamiltonian on cylindrical surface}
\label{sec:AppA}

The hexagonal cut of [111]-grown NWs appears due to presence of
$C_3$ symmetry axis. If we consider the cell of III-V semiconductor
(a cube in Fig.~\ref{fig1}b) then the projections of vertices of
this cube onto $(111)$-plane form a hexagon. The coordinate of
vectors from hexagon center to these vertices (see Fig.~\ref{fig1}c)
and to the middle of hexagon edges can be easily found from
geometric consideration. Introducing the auxiliary planar frame of
reference $\xi0\zeta$ (lying in $(111)$-plane), and choosing for the
definiteness ${\bf e}_\xi=\mathbf{d}_1=\frac{1}{\sqrt 2}(0,-1,1)$
and ${\bf e}_\zeta=\mathbf{a}_2=\frac{1}{\sqrt 6}(2,-1,-1)$ we find
that the coordinates of any unit vector lying in $(111)$-plane will
be determined by one parameter (azimuthal angle $\varphi$). We use
here simple model supposing the NW cross-section to be circular and
neglecting the hexagonal cut.

The cylindrical unit vectors have coordinates
$\mathbf{e}_r=\cos\varphi\mathbf{e}_\xi+\sin\varphi\mathbf{e}_\zeta$,
$\mathbf{e}_\varphi=-\sin\varphi\mathbf{e}_\xi+\cos\varphi\mathbf{e}_\zeta$,
$\mathbf{e}_z=\frac{1}{\sqrt 3}(1,1,1)$. Finally, we find for the
vectors $\mathbf{e}_r$ and $\mathbf{e}_\varphi$
\begin{equation}
{\bf e}_r=\sqrt{\frac
23}\left(\sin\varphi,-\cos\left(\varphi-\frac{\pi}{6}\right),\cos\left(\varphi+\frac{\pi}{6}\right)\right),
\end{equation}
\begin{equation}
{\bf e}_\varphi=\sqrt{\frac
23}\left(\cos\varphi,\sin\left(\varphi-\frac{\pi}{6}\right),-\sin\left(\varphi+\frac{\pi}{6}\right)\right).
\end{equation}

The straightforward procedure of derivation of $k$-linear DSOC
Hamiltonian from Hamiltonian (\ref{cubic}) in curvilinear
coordinates is quite difficult problem. We propose here more simple
approach: to construct the Hamiltonian for planar structure after
that to apply it to non-planar surface taking into account not too
large curvature.

Now we derive the $k$-linear DSOC Hamiltonian for planar
two-dimensional (2D) structure grown along the axis lying arbitrary
in plane (111). Consider planar 2D-structure that grown along vector
$\mathbf{e}_{x'}\equiv\mathbf{e}_r$. Axis $z'$ coincides with
(111)-direction and $y'$ is directed along $\mathbf{e}_\varphi$.

The coordinate transformation between crystallic coordinate system
($xyz$) and $x'y'z'$ reference frame performs by the following
matrix
\begin{equation}
A=\frac{1}{\sqrt{3}}\left(\begin{array}{ccc}
\sqrt 2\sin\varphi &\sqrt 2\cos\varphi & 1\\
-\sqrt 2\cos\left(\varphi-\frac{\pi}{6}\right) & \sqrt 2\sin\left(\varphi-\frac{\pi}{6}\right)& 1\\
\sqrt 2\cos\left(\varphi+\frac{\pi}{6}\right)&-\sqrt
2\sin\left(\varphi+\frac{\pi}{6}\right)&1
\end{array}\right).
\end{equation}

Usually at any rotation the spin matrix have to be transformed by
means on finite rotations matrices. However, for the case of spin
$1/2$ this transformation is simplified and the Pauli matrices
transform as components of usual vector \cite{Varshalovich1988}.
Applying the last transformation to spin matrices $\sigma_i$ and to
components of wavevector $k_j$ of (\ref{cubic}) after tedious
algebra and quantization of transformed $k^3$-Hamiltonian to 2D one
\cite{Dyakonov1986} we find (here we omit primes in coordinate
subscripts)
\begin{eqnarray}
\label{H_D_planar} \nonumber H_\mathrm{D}=\frac{\beta}{\sqrt
6}\bigg(-\frac{1}{\sqrt 2}\sigma_xk_y+\cos
3\varphi\sigma_xk_z\\
-\sin3\varphi\sigma_yk_z+3\sin 3\varphi\sigma_zk_y \bigg),
\end{eqnarray}
where $\beta=\gamma\langle k_x^2\rangle\simeq \gamma(\pi/w)^2$ with
$w$ being the width of 2D layer and $\langle...\rangle$ denotes the
averaging on the ground state of transverse motion. Here we take
into account that $\langle k_x\rangle=\langle k^3_x\rangle=0$ and
neglect the remaining $k^3$-terms owing to inequality $\langle
k_x^2\rangle\gg k^2_{y(z)}$.

It should be noted, that the symmetry analysis lead to the
Hamiltonian having all 6 possible terms (3 spin matrices multiplied
by 2 components of wavevector) for the group $C_1$ (having the
trivial symmetry operation only), but an exact calculation gives
only 4 contributions. However, there is no contradiction, because 2
other terms can arise from high-power contribution in bulk
spin-splitted Hamiltonian, e.g. from $k^5$-splitting. Moreover, the
form of the Hamiltonian (\ref{H_D_planar}) depends on the origin of
angle $\varphi$. Such a form take place when we suppose $\varphi=0$
to be corresponding to orientation along
$[0\overline{1}1]$-direction.

Assuming the TEG radius $r_0$ to be much larger than the lattice
parameter we obtain the Hamiltonian on the cylindrical surface from
planar one (\ref{H_D_planar}) by means of the following substitution

\begin{equation}
k_y\rightarrow
k_\varphi,\qquad\sigma_y\rightarrow\sigma_\varphi(\varphi),\qquad
\sigma_x\rightarrow\sigma_r(\varphi).
\end{equation}

Since in cylindrical coordinates some of Pauli matrices do not
commute with the operator $k_\varphi$ therefore we have to
additionally symmetrize (i.e. $AB\rightarrow \{A,B\}=(AB+BA)/2$)
obtained Hamiltonian in order to ensure the  Hermitiancy. After that
we obtain

\begin{eqnarray}
\nonumber H_\mathrm{D}=\frac{\beta}{\sqrt 6}\bigg[-\frac{1}{\sqrt
2}\{\sigma_r(\varphi),k_\varphi\}+\cos
3\varphi\sigma_r(\varphi)k_z\\
-\sin
3\varphi\sigma_\varphi(\varphi)k_z+3\sigma_z\{\sin3\varphi,
k_\varphi\}\bigg].
\end{eqnarray}

If we take into account the explicit action of operators in
symmetrized product then we obtain the final form of Hamiltonian
(\ref{H_D}).

\section{Matrix elements of Dresselhaus Hamiltonian}
\label{sec:AppB}

For the sake of simplicity now we decompose the Hamiltonian
(\ref{H_D}) into 3 terms,
$H_\mathrm{D}=H_\mathrm{D1}+H_\mathrm{D2}+H_\mathrm{D3}$, where

\begin{equation}
H_\mathrm{D1}=-\frac{\beta}{\sqrt{12}}\left(\sigma_r(\varphi)K_\varphi-\frac{i}{2r_0}\sigma_\varphi(\varphi)\right),
\end{equation}

\begin{equation}
H_\mathrm{D2}=\frac{\beta}{\sqrt 6}\sigma_r(-2\varphi)k_z,
\end{equation}

\begin{equation}
H_\mathrm{D3}=\sqrt{\frac32}\beta\sigma_z\left(\sin3\varphi
K_\varphi-\frac{3i}{2r_0}\cos 3\varphi\right).
\end{equation}

The matrix elements of these operators calculated on eigenfunctions
of Hamiltonian $H_0$ (\ref{eigenstates}) are given by

\begin{eqnarray}
\nonumber \langle
j's'k'|H_\mathrm{D1}|jsk\rangle=-\frac{\beta}{\sqrt{12}r_0}\delta_{kk'}\delta_{jj'}\lambda_j\\
\times\left(C^{(s')*}_{jk}D^{(s)}_{jk}+D^{(s')*}_{jk}C^{(s)}_{jk}
\right),
\end{eqnarray}

\begin{eqnarray}
\nonumber \langle j's'k'|H_\mathrm{D2}|jsk\rangle=\frac{\beta}{\sqrt
6}k\delta_{kk'}\\
\times\left(\delta_{j,j'-3}C^{(s')*}_{j'k}D^{(s)}_{jk}+\delta_{j,j'+3}D^{(s')*}_{j'k}C^{(s)}_{jk}\right),
\end{eqnarray}

\begin{eqnarray}
\nonumber
\langle
j's'k'|H_\mathrm{D3}|jsk\rangle=\sqrt{\frac32}\beta\frac{\delta_{kk'}}{2ir_0}\\
\nonumber \times \left\{
\delta_{j,j'-3}\left[(\lambda_j+1)C^{(s')*}_{j'k}C^{(s)}_{jk}-(\lambda_j+2)D^{(s')*}_{j'k}D^{(s)}_{jk}\right]\right.\\
-\left.\delta_{j,j'+3}\left[(\lambda_j-2)C^{(s')*}_{j'k}C^{(s)}_{jk}-(\lambda_j-1)D^{(s')*}_{j'k}D^{(s)}_{jk}\right]\right\}.
\end{eqnarray}

The concrete form of any matrix elements can be found using the
exact form of spinor components $C^{(s)}_{jk}$, $D^{(s)}_{jk}$ (see
Eq.(\ref{spinors})).


\end{document}